\documentstyle[12pt,aasms4]{article}
\tighten

\newcommand\be{\begin{equation}}
\newcommand\ee{\end{equation}}
\def\elll{l}
\def\dd{{\rm d}}

\input epsf
\lefthead{Basu et al.}
\righthead{Effect of peak asymmetry on structure of solar core}
\begin{document}

\title{Structure of the solar core: Effect of asymmetry of peak
profiles}

\author{S. Basu\altaffilmark{1,6},
S. Turck-Chi\`eze\altaffilmark{2},
G. Berthomieu\altaffilmark{3},
A. S. Brun\altaffilmark{2, 7},
T. Corbard\altaffilmark{3,8}, G. Gonczi\altaffilmark{3},
J.~Christensen-Dalsgaard\altaffilmark{4},
J. Provost\altaffilmark{3},
S. Thiery\altaffilmark{5},
A. H. Gabriel\altaffilmark{5},
P. Boumier\altaffilmark{5}}

\altaffiltext{1}{Institute for Advanced Study, Olden Lane, Princeton 
NJ 08540, USA}
\altaffiltext{2}{DAPNIA/Service d'Astrophysique, CEA/Saclay, \\
{\null}\qquad F-91191 Gif sur Yvette, Cedex 01, France}
\altaffiltext{3}{D\'epartement CASSINI, CNRS UMR 6529, 06304 Nice Cedex 4,
France}
\altaffiltext{4}{Teoretisk Astrofysik Center, Danmarks Grundforskningsfond, 
and \\
{\null}\qquad Institut for Fysik og Astronomi, Aarhus Universitet, 
DK 8000 Aarhus C, Denmark }
\altaffiltext{5}{Institut d'Astrophysique Spatiale, Orsay, France}
\altaffiltext{6}{Astronomy Department, Yale University, P.O. Box 208101, New
Haven, CT 06520-8101, USA}
\altaffiltext{7}{JILA, University of Colorado, Boulder, CO 80309-0440, USA}
\altaffiltext{8}{High Altitude Observatory Boulder, CO 80301, USA}

\begin{abstract}
Recent studies have established that peaks in solar oscillation power
spectra are not Lorentzian in shape, but have a distinct asymmetry.
Fitting a symmetric Lorentzian profile to the peaks therefore produces
a shift in frequency of the modes. Accurate determination 
of low-frequency modes is essential to infer the structure of the
solar core by inversion of the mode frequencies. In this paper we
investigate how the changes in frequencies of low-degree modes obtained by
fitting symmetric and asymmetric peak profiles change the inferred
properties of the solar core. We use data obtained
by the Global Oscillations at Low Frequencies (GOLF) project on board the
SoHO spacecraft. Two different solar models and  inversion 
procedures are used
to invert the data to determine the sound speed in the solar core.
We find that for a given set of modes 
no significant difference in the inferred sound-speed results
from taking asymmetry into account when fitting the low-degree modes.
\end{abstract}

\keywords{Sun: oscillations; Sun: interior}

\section{Introduction}

Accurate determination of the frequencies of low-degree oscillation
of the Sun is essential in inferring the structure of the solar core,
owing to the
small effect of this part of the Sun on the oscillation frequencies
(e.g. Turck-Chi\`eze, Brun, \& Garcia 1999).
Thus particular care is required in interpreting the data obtained.
Important aspects, which have received considerable attention recently,
are the effect on the frequencies of the variations of solar activity during
the solar cycle (Dziembowski \& Goode 1997; Dziembowski et al. 1997),
as well as the different results obtained from intensity and
velocity measurements (Toutain et al. 1997).
These different observations have led
to a better understanding of how the data must be analyzed,
emphasizing, for example, the need to use
the same period of time for low- and high-degree modes
to avoid differences in the near-surface effects,
and ideally considering data sets of one or a maximum of two years
around the minimum of activity.

There is now a number of experiments designed specifically to
investigate the properties of the solar core; one
of these is the Global Oscillations at Low Frequencies (GOLF)
instrument (e.g. Gabriel et al.~1997)
on board the Solar and Heliospheric Observatory (SoHO)
(e.g. Domingo, Fleck \& Poland 1995).
A comparison of different data series still shows differences in the inferences
in the solar core substantially exceeding the error bars
(e.g. Figure 6 of Turck-Chi\`eze et al. 1997).
In order to make progress on this point, we propose here a
differential study, using GOLF data taken near the solar minimum, to show the
influence of three ingredients of the investigation:
(a) the asymmetry of the frequency peaks due to the localized nature
of the source of the oscillation and its interaction with the noise;
(b) the role of the methods used for inverting the data; and
(c) the influence of the solar model used
(with different prescriptions for the atmosphere) to perform the inversion.
The purpose of the paper is to give quantitative
estimates of these effects and  a better estimate of the uncertainty
they may introduce in our determination of the structure of the solar core.

It has been
demonstrated that in general the peaks in solar
oscillation power spectra are not
symmetric (e.g. Duvall et al.~1993;  Nigam \& Kosovichev 1998;
Toutain et al.~1998; Chaplin \& Appourchaux 1999).
This is believed to be a consequence of the
localized nature of the source which drives the oscillations
(e.g. Gabriel 1993; Abrams \& Kumar 1996; Roxburgh \& Vorontsov
1997; Nigam et al. 1998; Rosenthal 1998). Despite the
evidence for asymmetry, most analyses of observed solar power spectra
involve the fitting of symmetric Lorentzian profiles to the
peaks in power. This leads to a systematic error in the
inferred frequencies.

While there is no longer any doubt that the peaks are asymmetric
and that there is a frequency shift, what is still not clear
is whether the shift in the frequencies changes results obtained
by inverting the frequencies. Tests by Christensen-Dalsgaard 
et al. (1998) and Rabello-Soares et al. (1999b)
suggest that the frequency shift is a smooth function of frequency,
which may possibly be removed while inverting the frequencies.
Also, using frequencies determined from $m$-averaged spectra obtained
the Global Oscillation Network Group (GONG), Basu \& Antia (2000)
showed that frequencies obtained by  using asymmetric fits to the peaks do not
significantly change results obtained by inversion.
However, Toutain et al. (1998) found that for low-degree data obtained
by the Michelson Doppler Imager (MDI) on board SoHO for 679 days
of observation, the shift in the frequencies does change the result
substantially in the solar core.

In this paper, we use solar data obtained by GOLF to check whether
there is a significant change in the inferred solar structure
when one shifts from using frequencies from Lorentzian fitted peaks to 
those
from asymmetric profiles. The details of the fitting procedure and
the results of the fits have been described by Thiery et al. (1999).
We use only the $\elll=0$, 1 and 2 modes from  the GOLF data. Since
it is essential to have intermediate- and high-degree modes to do a
reliable inversion, we use the $\elll=$ 3-250 data
obtained from MDI
(Schou et al. 1998).
We note that only symmetrically fitted frequencies are available
for these observations.
This introduces an unfortunate inconsistency in our analysis which,
however, would most likely increase the error in the inferred
sound speed (e.g. Rabello-Soares et al. 1999b).

The rest of the paper is organized as follows: we describe the 
inversion techniques
and solar models used in this work in \S 2, the results of the 
inversions are
discussed in \S 3 and our conclusions are stated in \S 4.

\section{The inversion technique}

Inversions to determine solar structure from solar oscillation frequencies
proceed through the linearization of the equation for linear
adiabatic oscillations around a known solar model.
When the oscillation equation is linearized --- under the assumption
of hydrostatic equilibrium --- the fractional change in the
frequency can be related to the fractional changes in the squared
sound speed ($c^2$) and density ($\rho$).  Thus,
\begin{equation}
\frac{\delta\omega_i}{\omega_i}  \! = \!
\int K_{c^2,\rho}^i (r) \frac{\delta
c^2}{c^2}(r) \dd r
+ \int K_{\rho,c^2}^i (r) \frac{\delta \rho}{\rho}(r) \dd r
 + \frac{F_{\rm surf}(\omega_i)}{E_i}\; 
\label{eqn:freqdif}
\end{equation}
(cf. Dziembowski et al. 1990).  Here $\delta \omega_i$ is the
difference in the frequency $\omega_i$ of the $i$th mode between the
solar data and a reference model.
The kernels $K_{c^2,\rho}^i$
and $K_{\rho,c^2}^i$ are known functions of the reference model which
relate the changes in frequency to the changes in $c^2$ and $\rho$,
respectively; and $E_i$ is the inertia of the mode, normalized by the
photospheric amplitude of the displacement.  The term $F_{\rm surf}$
results from the near-surface  differences between the Sun and models 
because of the 
difficulty in modeling the outer layers.

\subsection{The inversion methods}

We have used two different methods to invert the frequencies: the 
Regularised
Least Squares (RLS) and the Subtractive Optimally Localized Averages
(SOLA) methods.

For RLS inversions, the sound-speed and density differences between the
Sun and the reference model are described by a set of basic functions
in radius, $r$ (in this case  splines). The surface term is described
as a spline in frequency. The spline coefficients are
found by minimizing the
difference between the left-hand side of Eq.~(\ref{eqn:freqdif}) and
the right-hand side expanded in
splines, subject to the condition that the resulting $\delta c^2/c^2$ is
smooth.

The number of knots in radius $r$ (a total of 120) and the number of knots
in frequency
to describe the surface term (25) are determined by the fact that 
for a proper inversion we need enough knots to ensure that  
the residuals of the fit are randomly distributed as a function of frequency 
and
lower turning point of the modes for a proper inversion;
on the other hand, the 
condition number of the system of equations (which increases with 
an increase in the numbers of knots) should be as small as possible to
ensure that the system is sufficiently well-conditioned to
allow a stable numerical solution
(see e.g. Basu \& Thompson 1996).
The knots were
distributed according  to the density  of turning points of the
 set of modes along the radius.
The trade-off parameter which controls the error in the solution and its
smoothness was determined by plotting the
 so-called L-curve  which gives the Tikhonov smoothing term
(here the norm of the first derivative of the solution) as a
 function of $\chi^2$ to find a compromise between a good fit of
the data (small $\chi^2$) and a rather smooth physically acceptable 
solution
(Gonczi et al. 1998).

The principle of the SOLA inversion technique (Pijpers \& Thompson 1992)
is to form linear combinations of Eq.~(\ref{eqn:freqdif}) with
weights $d_i(r_0)$ chosen so as to obtain an average  of
$\delta c^2/c^2$  localized near
$r=r_0$ while suppressing the contributions from
$\delta\rho/\rho$,
and the near-surface errors,
when inverting for $\delta c^2/c^2$.
In addition, the statistical errors in the combination must be
constrained.
The result of the inversion is then an average of $\delta c^2/c^2$, 
with a weight determined by the {\it averaging kernel}, defined by
\be
{\cal K}(r_0,r)=\sum d_i(r_0)K^i_{c^2,\rho}(r) 
\label{eqn:avker}
\ee
and normalized so that $\int {\cal K}(r_0,r) \dd r =1$.
%and is the area over which $\delta c^2/c^2$ is averaged.
Details of the implementation were provided by
Basu et al. (1996) and a procedure to find the parameters required
in the inversion was
discussed by Rabello-Soares, Basu \& Christensen-Dalsgaard  (1999a).

\subsection{The reference solar models}

We have used two reference models for this work.
The first model (hereafter referred to as the ``Saclay/Nice model'')
is an updated calculation of the standard model
of Brun, Turck-Chi\`eze \& Morel (1998) based on the
CESAM code (Morel 1997). Nuclear reaction rates of
Adelberger et al. (1998) were used with
screening effects from Dzitko et al. (1995). The most recent
OPAL opacity tables (Iglesias \& Rogers 1996) and the OPAL
equation of state (Rogers, Swenson \& Iglesias 1996) have been introduced
in constructing the model.  This model converged at the solar age with
 the observed
abundances of  thirteen elements from Grevesse \& Noels (1993)
and microscopic diffusion of each of these elements was
computed using diffusion coefficients from Michaud \& Proffitt (1993).
A reconstructed atmosphere was
deduced from the ATLAS9 atmosphere code of Kurucz (1991).
The computation included a pre-main-sequence evolution phase
and the model has an age of 4.6 Gyr, including this phase. A more detailed
description of the
model was given by Brun et al. (1998).
Some results of comparisons of this model with others and
preliminary comparisons with the Sun were made by
Turck-Chi\`eze et al. (1998).

The second model
is Model S of Christensen-Dalsgaard et al. (1996). This model is
used because many helioseismological results in literature are
based on this reference model.
This is a standard solar model constructed with
the Livermore (OPAL) equation of state (Rogers et al. 1996).
For
temperatures higher than $10^4$ K, an early version of the OPAL opacities 
was used (Rogers \& Iglesias 1992);
at lower temperatures, opacities from the tables
of Kurucz (1991) were taken. The model incorporates the diffusion of
helium and heavy elements below the convection zone. The surface
heavy element ratio is $Z/X = 0.0245$ (Grevesse \& Noels 1993).  The
model has an age of 4.6 Gyr without pre-main-sequence. The model was
described in detail
by Christensen-Dalsgaard et al. (1996).

\section{Results}

\begin{figure}
%\plotone{fig1.eps}
\plotfiddle{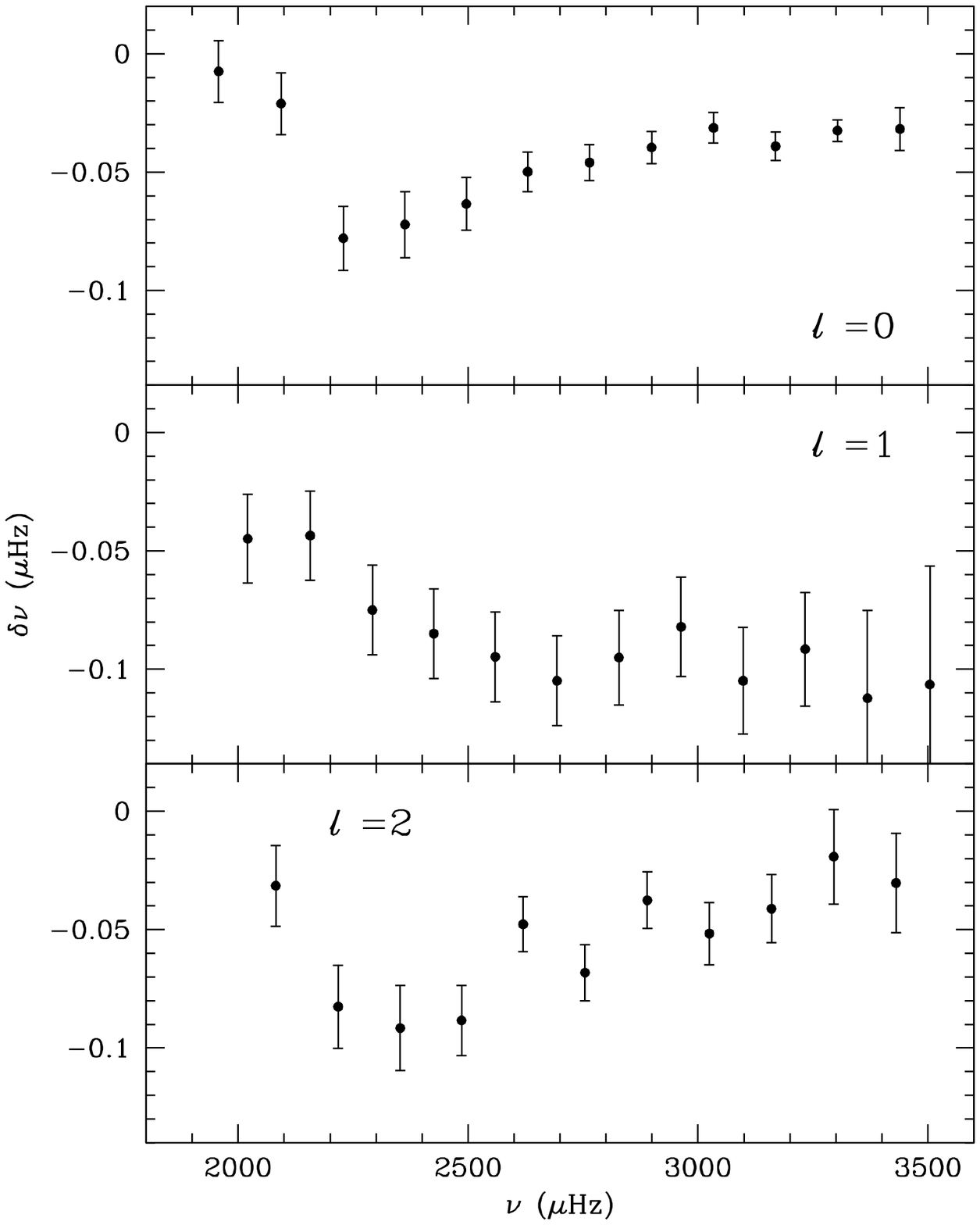}{13 true cm}{0}{80}{80}{-255}{-140}
\figcaption{The difference between the frequencies obtained by 
fitting a Lorentzian profile to the peaks in the oscillation 
power spectrum and those obtained by fitting  an asymmetric profile,
in the sense (symmetric fit) -- (asymmetric fit).
The differences are shown only for the frequency range used in the
inversions. 
\label{fig:freqdif}
 }
\end{figure}

The differences in frequencies obtained by the fitting Lorentzian
profiles and those obtained by fitting an asymmetric profile are
shown in Fig.~\ref{fig:freqdif}. Note that the differences
are systematic, not random, and hence one could expect 
that the result of inverting the two sets of frequencies will
show systematic differences too.

\begin{figure}
\plotone{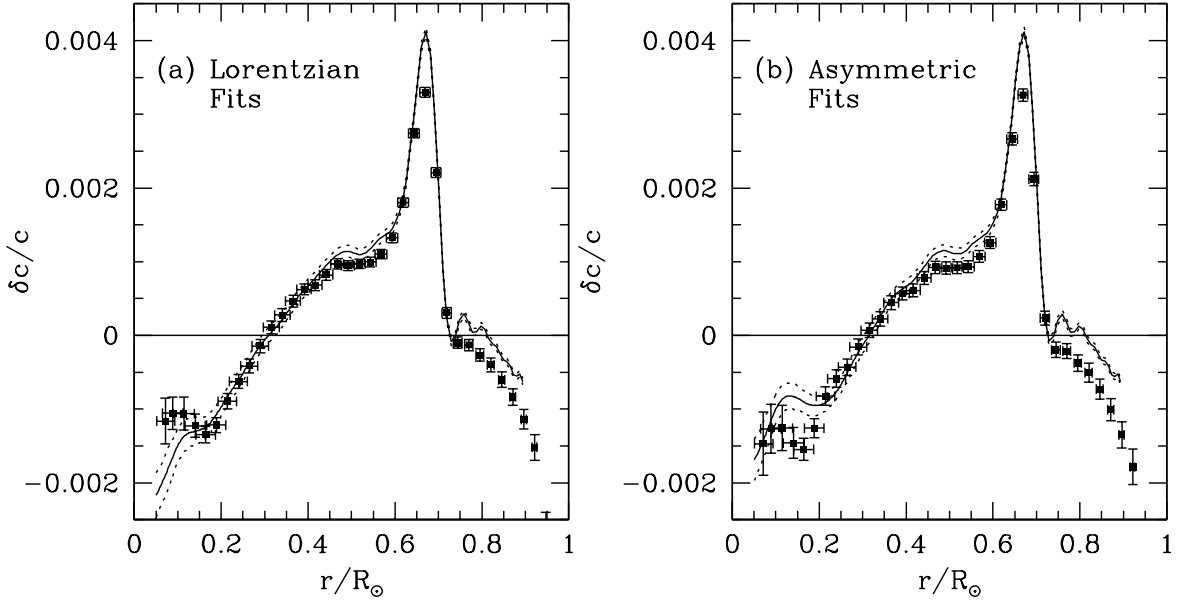}
\figcaption{ The relative sound-speed differences, $\delta c/c$
 between the Sun and the Saclay/Nice model.
The differences are taken in the sense (Sun-model)/Sun.
Panel (a) is the results of inverting frequencies obtained by
fitting power spectrum peaks with a Lorentzian profile. Panel (b) is
from frequencies obtained by fitting an asymmetric profile
to the peaks. In both panels data for models with $\elll=0$, 1 and 2
were obtained
from the GOLF instruments. Data for modes with higher degree
were obtained by the MDI instrument. The points with error-bars
are the results of the SOLA inversions. The vertical error bars
indicate the 1$\sigma$ error in the inversion due to data errors
while the horizontal error bars are a measure of the resolution.
The solid line is the result of an RLS inversion, and the dotted lines
show the 1$\sigma$ error limits on that.
\label{fig:saclay}
 }
\end{figure}

Figure \ref{fig:saclay} shows the sound-speed difference between the 
Saclay/Nice model
and the
Sun obtained using GOLF data combined with MDI data for $\elll \ge 3$.
Both sets of GOLF frequencies, i.e., the one obtained by fitting a
Lorentzian profile and the one obtained by fitting the
asymmetric profile of Thiery et al. (1999), are shown.

The results of the SOLA and RLS inversions agree within errors 
in most of the Sun.
The height of the bump at the base of the convection zone is higher in the
case of RLS than SOLA and is a result of slightly different resolutions
in the two inversions. For the same error magnification, RLS generally
has a better resolution than SOLA, though with the drawback that the
RLS averaging kernels have some structure far away from the target
radius.
There is some difference in the
convection zone, which is most probably a reflection of differences
in error correlation.

\begin{figure}
\plotfiddle{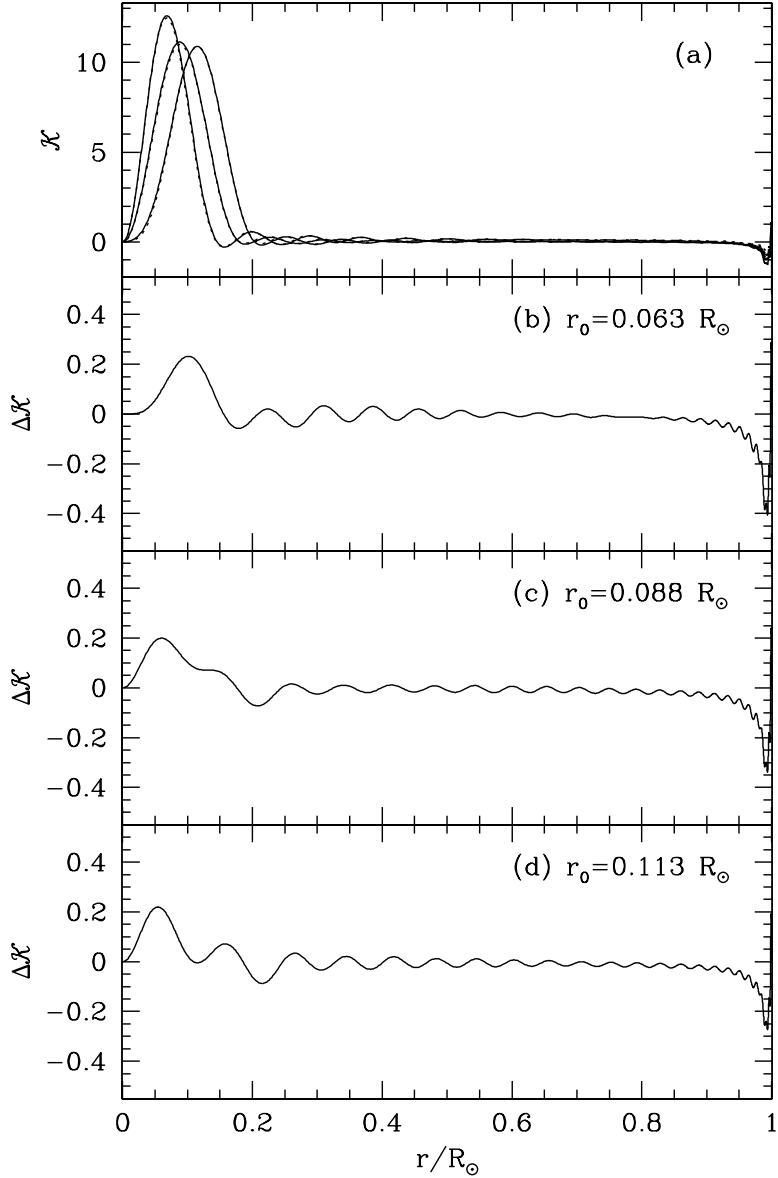}{13 true cm}{0}{80}{80}{-255}{-140}
\figcaption{(a) A sample of SOLA
averaging kernels. The target radii
for the kernels are 0.063 $R_\odot$, 0.088 $R_\odot$ and
$0.113 R_\odot$. The continuous lines are for the mode-set obtained
by fitting a Lorentzian, the dotted lines --- which can barely be
distinguished from the continuous lines --- are for the
asymmetric set. Panels (b)-(d) show the differences,
in the sense (Lorentzian $-$ asymmetric),
between the
averaging kernels of the two sets. Note that the differences
are much smaller than the peak height of the averaging kernels.
\label{fig:avkergolf}
}
\end{figure}

For a given inversion method, the results for the outer layers
are almost identical regardless of the data set used.
This is expected since the same set of high-degree modes is
used in both cases. There are changes in the core when the
symmetric data set is replaced by the asymmetric set; however,
we can see that they are % not significant and 
within the errors, and hence not significant.
In Fig.~\ref{fig:avkergolf} we show
some of the averaging kernels in the region of the core.
The figure also shows the difference between the
averaging kernels obtained for the inversion of the two
data sets. We see that there is very little difference
between them.

\begin{figure}
\plotone{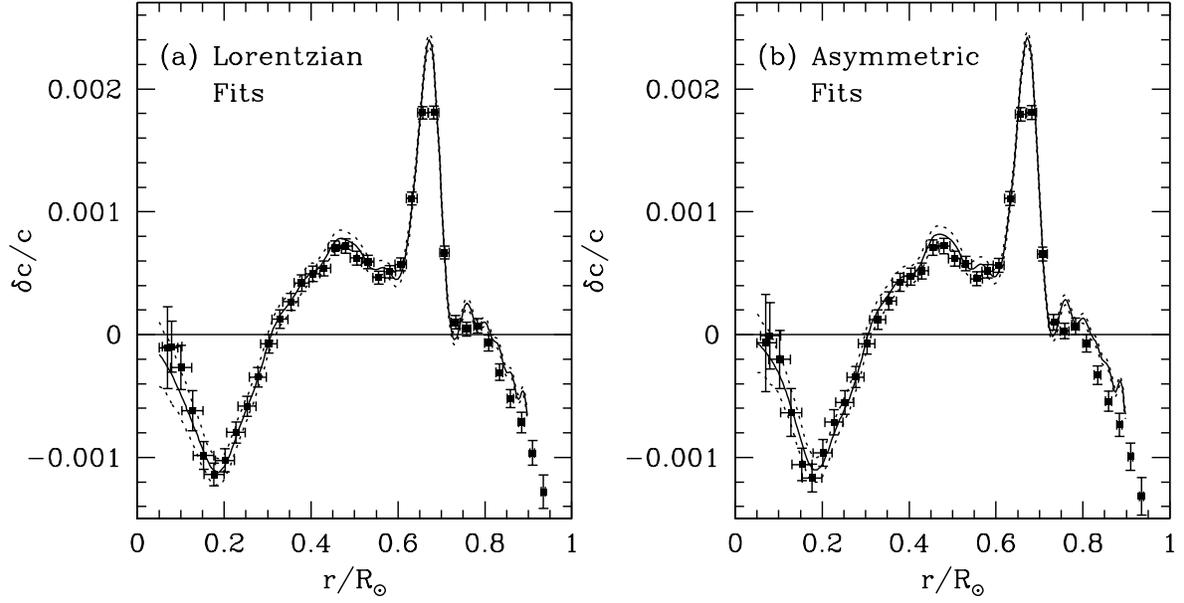}
\figcaption{The relative sound-speed differences, $\delta c/c$
 between the Sun and Model S using GOLF and MDI data.
The differences are taken in the sense (Sun $-$ model)/Sun.
Panel (a) is the results of inverting frequencies obtained by
fitting power spectrum peaks with a Lorentzian profile. Panel (b) is
from frequencies obtained by fitting an asymmetric profile
to the peaks.
\label{fig:modelS}
}
\end{figure}

We get similar conclusions using Model S (cf. Fig.~\ref{fig:modelS}):
the introduction of the asymmetry does not affect significantly the inversion 
in the core.
The results for the outer layers are identical, while the change
in the core is very small when data sets are changed.
The two models, however, have  different  sound-speed profiles which
warrant some comments. The sound-speed differences between the two models
can be attributed completely to the physical inputs of the models.
The differences are mainly  due to the following three
effects: (1) a reestimate of the solar age
(about 4.55 Gyr without including the pre-main-sequence)
in the Saclay/Nice model which lowers the relative sound-speed difference
in the very inner core
and slightly increases the peak in the sound-speed difference
relative to the Sun below the base of the
convection zone, (2) a reestimate of the nuclear reaction
rates and the screening effect which has similar consequences
(see also the discussion of Turck-Chi\`eze et al. 1998),
and (3) the upgrade to the most recent OPAL opacity tables 
which has little effect in the core but dominates the increase in
the sound-speed difference below the convection zone.

Our main result is 
that frequencies obtained using the asymmetric fits
do not change inversion results.
This is consistent with the estimates by Christensen-Dalsgaard et al. (1998) 
of the functional form of frequency shifts caused by asymmetry,
based on artificial data,
which indicated that such shifts would largely be eliminated
together with the surface term in $F_{\rm surf}$.
This conclusion was confirmed by
the inverse analyses carried out by Rabello-Soares et al. (1999b)
of such artificial data which showed that asymmetric fits had a very
modest effect on the results of structure inversion.
However, the results are strikingly different from those
obtained by Toutain et al. (1998) using only MDI data.

\begin{figure}
\plotone{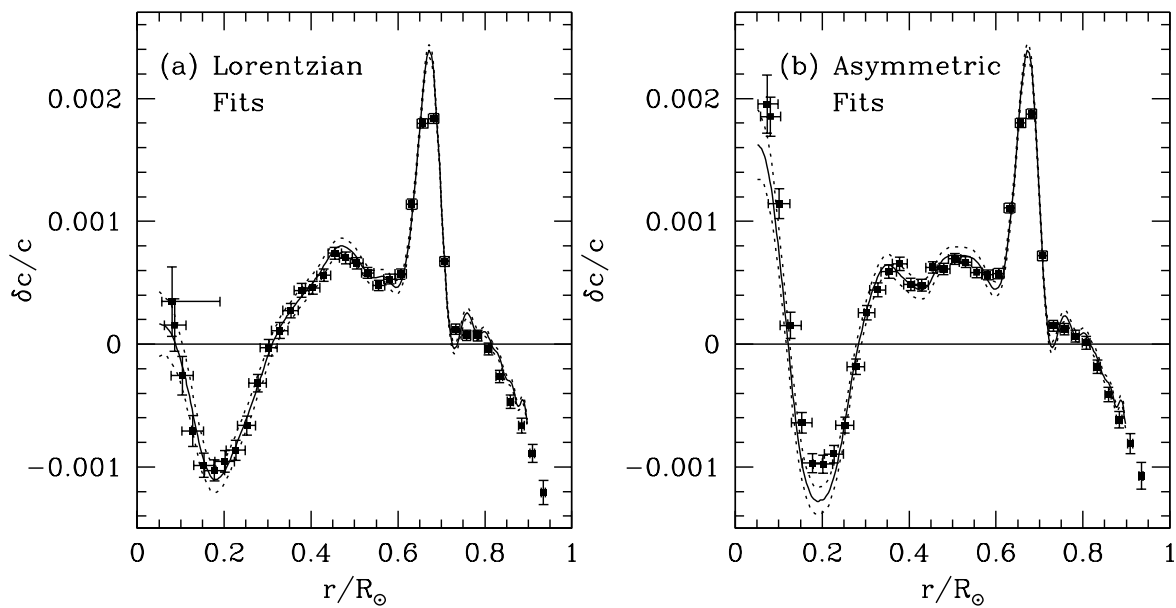}
\figcaption{The relative sound-speed differences, $\delta c/c$
 between Model S and the Sun using MDI data only.
Panel (a) shows the result of using frequencies obtained
by fitting symmetric profiles. Panel (b) uses the
$\elll=$ 0, 1 and 2 data from Toutain et al. (1998).
Higher-degree frequencies are from the MDI set.
\label{fig:mdi}
}
\end{figure}

The question then arises as to why the MDI results of Toutain et al.
are so discrepant. In Fig.~\ref{fig:mdi} we show the inversion of the 
symmetric
MDI  set (Schou et al. 1998), as well as the inversion of the low-degree MDI data
obtained by fitting an asymmetric profile (Toutain et al.  1998)
combined with the $\elll \ge 3$
data of the MDI data set;
in both cases the reference model was Model S.
We see that while the
symmetric MDI data give results quite similar to those based on GOLF,  the
asymmetric MDI data do indeed give quite different results.

\begin{figure}
\plotfiddle{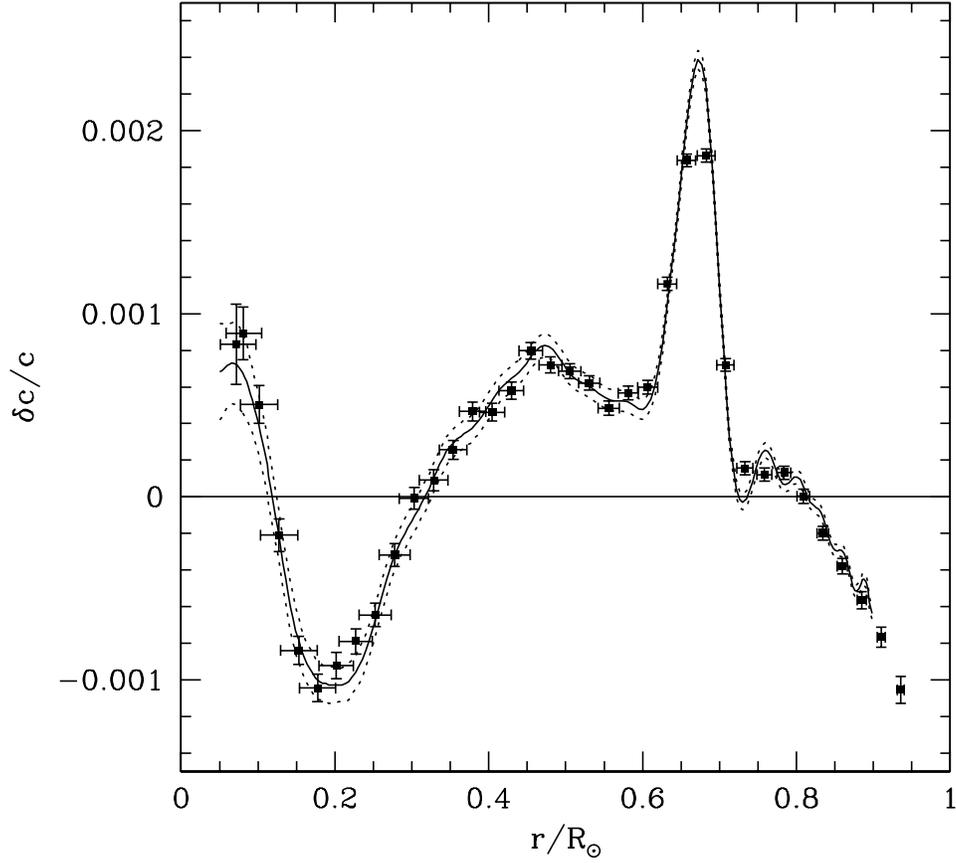}{10 true cm}{0}{90}{90}{-280}{-210}
\figcaption{The same as Fig.~\ref{fig:mdi}(b) but for modes
$\elll=2$ $n=6$ and $\elll=2$ and $n=7$ removed from the set.
Note that the sound-speed difference in the core has
decreased compared with the result in Fig.~\ref{fig:mdi}(b).
%Also the result of the outer layers is now identical.
\label{fig:mdib}
}
\end{figure}

The MDI asymmetric set has some modes with much lower errors than their
adjacent modes. These modes thus get very large weights in the
inversion process and, since there are very few low-degree modes anyway,
they can indubitably influence the inversion result.
This does indeed seem to be the case.  
Furthermore, 
the modes $\elll=2,n=6$ and $\elll=2,n=7$ are suspect because of
the fact that they have extremely large
residual ($> 10\sigma$) in the RLS inversions.
Removal of these modes
reduces the difference in the results between the symmetric and
asymmetric MDI sets (cf. Fig.~\ref{fig:mdib}).
Therefore we find no evidence 
even from the thus corrected MDI data that
frequencies obtained with asymmetric profiles fitted to the
peaks in the power spectrum  cause significantly different
results, compared with frequencies obtained with a Lorentzian fit.

\begin{figure}
\plottwo{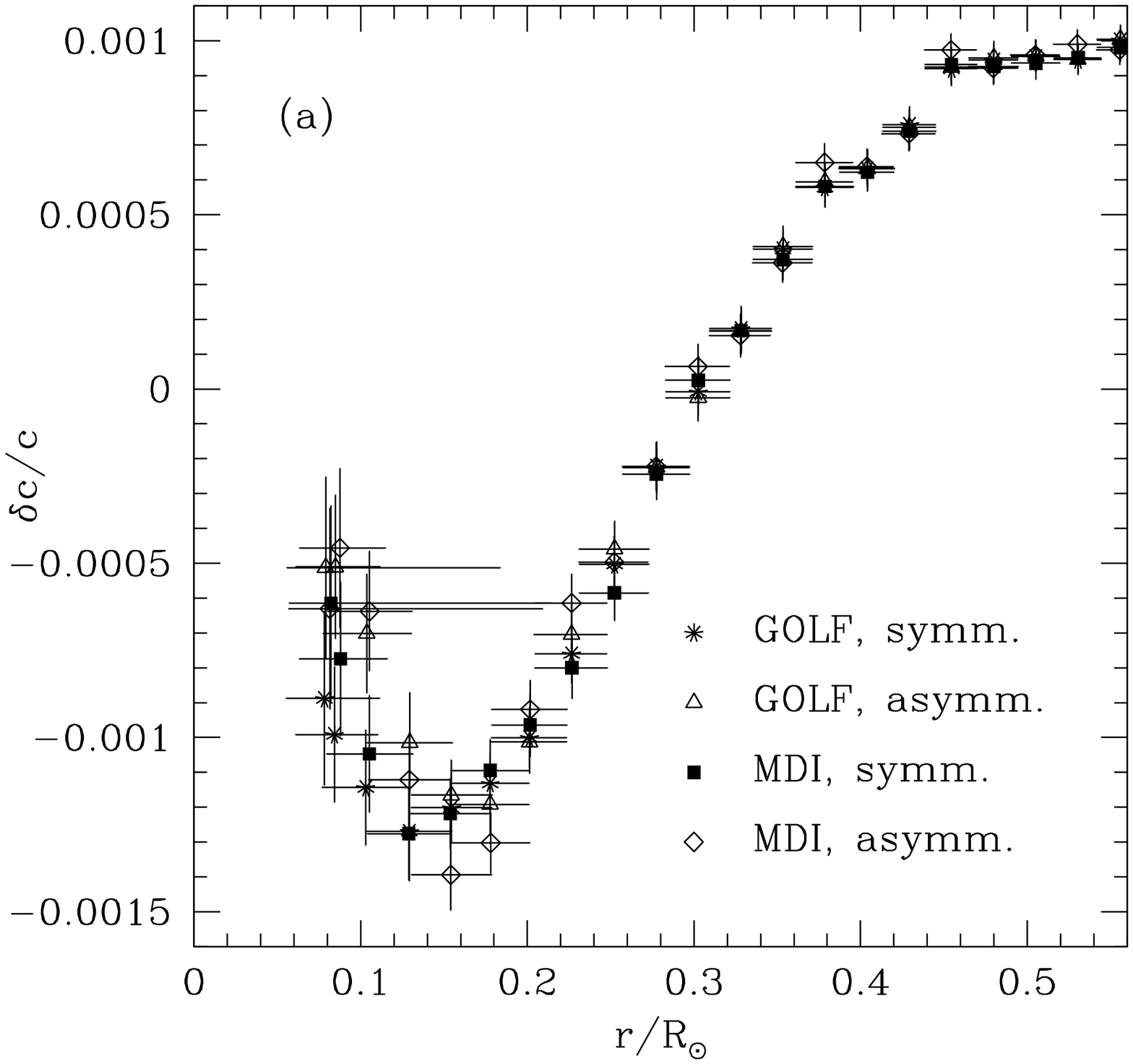}{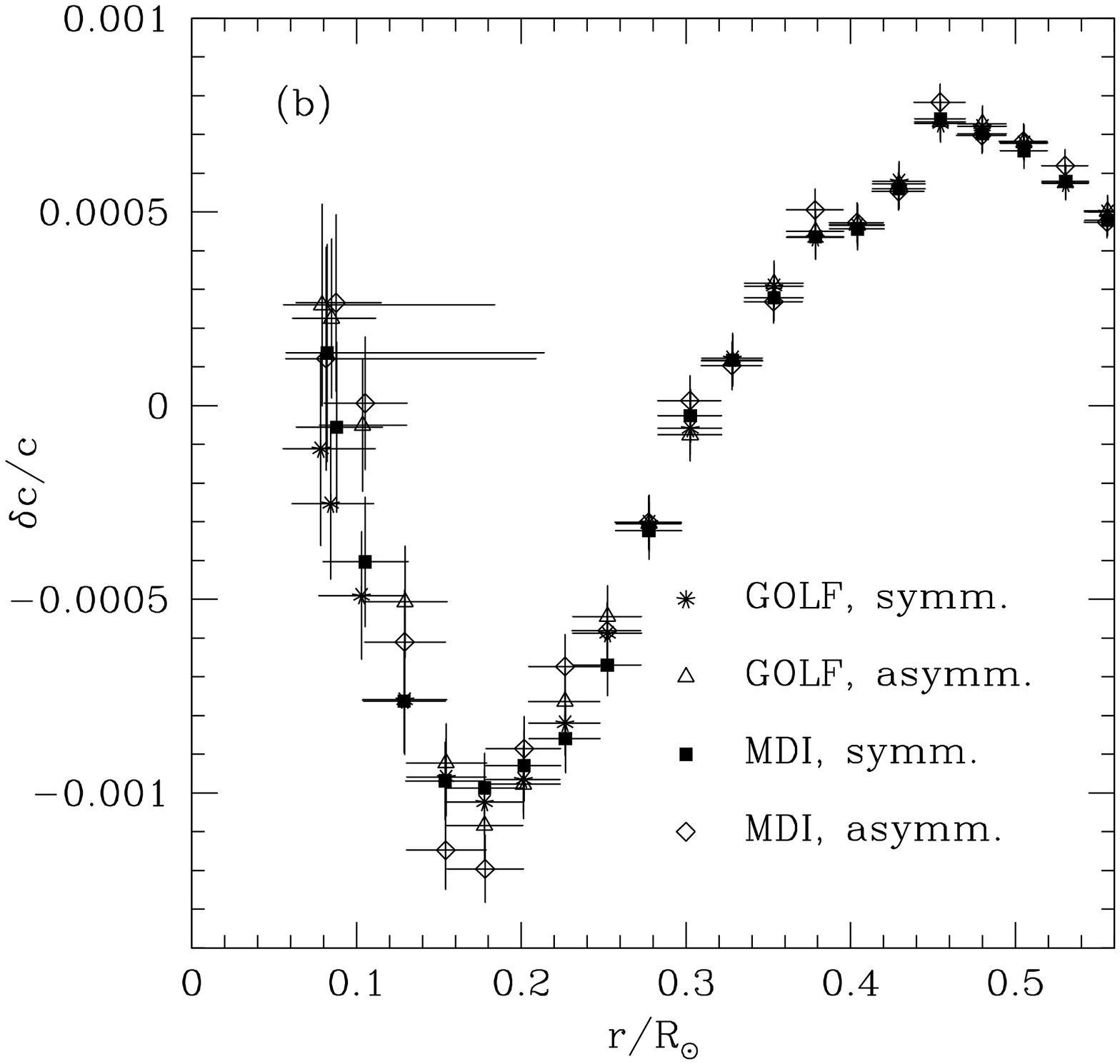}
\figcaption{The sound-speed difference between the Sun and the
Saclay/Nice model 
[Panel (a)] 
and the Sun and Model S 
[Panel (b)] as obtained 
by inverting a
common set of modes for the
4 data sets (i.e., GOLF symmetric and asymmetric; MDI
symmetric and asymmetric). The label ``symm.'' and ``asymm.''
applies only to modes of degree 0, 1 and 2. Higher-degree modes
in all four sets are from the MDI set and were obtained by fitting
a purely symmetric Lorentzian profile.
\label{fig:common}
}
\end{figure}

It may be noted here that the $l=1$ modes of the GOLF and MDI sets
show a fairly large difference at high frequency. However, the errors
on the modes are also very high and the difference does not seem to
cause substantial difference in the inversion results.
There is, however, some remaining difference between the GOLF inversion
results and the MDI inversion results. To check whether that is merely
an artifact of having different numbers of modes in the different
data sets, we have inverted a common set of modes from each data set.
The set has $l=0$ and $l=1$ modes  of $n=13$ to $n=25$, and  
$l=2$ modes with $n=13$ to $n=23$. The higher-degree modes are from 
the MDI set as before.
The inferred sound-speed differences in the inner parts of the Sun
are shown in Fig.~\ref{fig:common}. 
%Only the inner regions are shown.
Note that the results are actually quite similar for both reference
models.

\section{Conclusions}

We find very little evidence that the shift in frequencies between
symmetric and asymmetric fits to solar oscillation power
spectra changes our inferences concerning the sound speed in the solar core.
The changes we see are well within one standard deviation. Larger
differences occur as a result of addition of modes to the set, or
when using different inversion methods.
It should be kept in mind that this conclusion is based
on combining symmetrically or asymmetrically fitted low-degree
data with higher-degree frequencies obtained from symmetric fits.
This introduces an inconsistency in the results based on the asymmetrical fits,
which could be significant in view of the fact that 
the low-degree modes form a very small fraction of the total mode set.
On the other hand, an inconsistency of this kind, by introducing
a degree-dependent systematic error in the frequencies,
would appear likely if anything to increase the effect on the inversions.
We note also that Rabello-Soares et al. (1999b) found little effect
of asymmetry in inversions of artificial data including asymmetry.
Similarly, the analysis by Basu \& Antia (2000), including both low- and 
intermediate-degree asymmetrically fitted modes,
indicates that the results obtained in this
work will not change; however, it should be kept in mind that 
Basu \& Antia  obtained the frequencies
from $m$-averaged spectra and errors in
averaging may affect the results. 
Thus it is obvious that our study must be repeated when
asymmetric fits to peaks of intermediate- and high-degree
modes are also available. 
Of course, when this analysis will be generalized,
careful attention will be needed to other sources of distortion,
such as those resulting from the effect of the solar cycle on the
outer layers which will be considerable for data taken between
1998 and 2003.

This study confirms that there is a  dip in the sound-speed difference
at about 0.18 $R_\odot$, which appears independent of the model and 
inversion
procedure used.
Even deeper into the core,
the difference observed appears more dependent
on the details of a few modes of low order which
we can determine with high accuracy,
and on the use of symmetric or asymmetric profiles
in fitting the modes,
as well as on the surface treatment in the inversion.
In addition, the errors in the inversion results caused by
data errors are larger.
Knowledge about this region of the Sun,
crucial for the neutrino predictions, will be
improved by extended data series
and coherent treatment of asymmetric fitting of all the modes.

\acknowledgements

This work  utilizes data from GOLF and the Solar Oscillations
Investigation / Michelson Doppler Imager (SOI/MDI) on the Solar
and Heliospheric Observatory (SoHO).  SoHO is a project of
international cooperation between ESA and NASA.
The work was supported in part
by the Danish National Research
Foundation through its establishment of the Theoretical Astrophysics Center.

\clearpage

\end{document}